%% file: document.tex
\documentclass[sigconf]{acmart}

\usepackage{ctable}
\usepackage{multirow}
\usepackage{balance}
\usepackage{booktabs}
\usepackage{hyphsubst}
\usepackage{graphicx}
\usepackage{amsmath}
\usepackage{url}
\usepackage{breakurl}
\usepackage{hhline}
\usepackage[caption = false]{subfig}

\newcounter{lizcounter}
\DeclareRobustCommand{\liz}[1]{\textbf{/* #1 (liz) */}\stepcounter{lizcounter}\typeout{LaTeX Warning: liz comment \thelizcounter: #1 (line \the\inputlineno)}}

\newcounter{findingscounter}

\newcommand{\para}[1]{\vspace{2mm}\noindent\textbf{#1}}

\copyrightyear{2019} 
\acmYear{2019} 
\setcopyright{acmcopyright}
\acmConference{REVEAL@RecSys'2019}{September 16-20, 2019}{Copenhagen, Denmark}
\acmPrice{15.00}
\acmDOI{xx.xxx/xxxxxx.xxxxxx}
\acmISBN{xxx-x-xxxx-xxxx-x/xx/xx}

\begin{document}

\title[Evaluating Tripartite Recommendations in Data Markets]{Using the Open Meta Kaggle Dataset to Evaluate Tripartite
Recommendations in Data Markets}

\author{Dominik Kowald}
\affiliation{%
  \institution{Know-Center GmbH}
  \city{Graz, Austria} 
}
\email{dkowald@know-center.at}

\author{Matthias Traub}
\affiliation{%
  \institution{Know-Center GmbH}
  \city{Graz, Austria} 
}
\email{mtraub@know-center.at}

\author{Dieter Theiler}
\affiliation{%
  \institution{Know-Center GmbH}
  \city{Graz, Austria} 
}
\email{dtheiler@know-center.at}
\author{Heimo Gursch}
\affiliation{%
  \institution{Know-Center GmbH}
  \city{Graz, Austria} 
}
\email{hgursch@know-center.at}
\author{Emanuel Lacic}
\affiliation{%
  \institution{Know-Center GmbH}
  \city{Graz, Austria} 
}
\email{elacic@know-center.at}
\author{Stefanie Lindstaedt}
\affiliation{%
  \institution{Know-Center GmbH}
  \city{Graz, Austria} 
}
\email{slind@know-center.at}
\author{Roman Kern}
\affiliation{%
  \institution{Graz University of Technology}
  \city{Graz, Austria}  
}
\email{rkern@tugraz.at}
\author{Elisabeth Lex}
\affiliation{%
  \institution{Graz University of Technology}
  \city{Graz, Austria} 
}
\email{elisabeth.lex@tugraz.at}

\renewcommand{\shortauthors}{Kowald et al.}

\begin{abstract}
This work addresses the problem of providing and evaluating recommendations in data markets. Since most of the research in recommender systems is focused on the bipartite relationship between users and items (e.\,g., movies), we extend this view to the tripartite relationship between users, datasets and services, which is present in data markets. Between these entities, we identify four use cases for recommendations: (i) recommendation of datasets for users, (ii) recommendation of services for users, (iii) recommendation of services for datasets, and (iv) recommendation of datasets for services. Using the open Meta Kaggle dataset, we evaluate the recommendation accuracy of a popularity-based as well as a collaborative filtering-based algorithm for these four use cases and find that the recommendation accuracy strongly depends on the given use case. The presented work contributes to the tripartite recommendation problem in general and to the under-researched portfolio of evaluating recommender systems for data markets in particular.
\end{abstract}


\maketitle

\section{Introduction}
Data-driven services are becoming an increasingly important aspect of the modern economy, with data markets playing a key role as broker between the stakeholders of the data-driven ecosystem.
Various initiatives have been started to research the requirements and dynamics of data markets. To name two examples, the ``Data Market Austria'' (DMA)\footnote{\url{https://datamarket.at/en/}}~\cite{traub2017} is a national project in Austria, while ``A European AI On Demand Platform and Ecosystem'' (AI4EU)\footnote{\url{https://www.ai4eu.eu/}} aims at creating a market platform for data and artificial intelligence solutions on the European level.

For successful collaborations in data markets, the different entities need to collaborate with each other in order to create new solutions and to be able to provide innovative data products~\cite{Cavanillas2016,curry2016big}.

\begin{figure}[t!]
   \centering
	\includegraphics[width=0.45\textwidth]{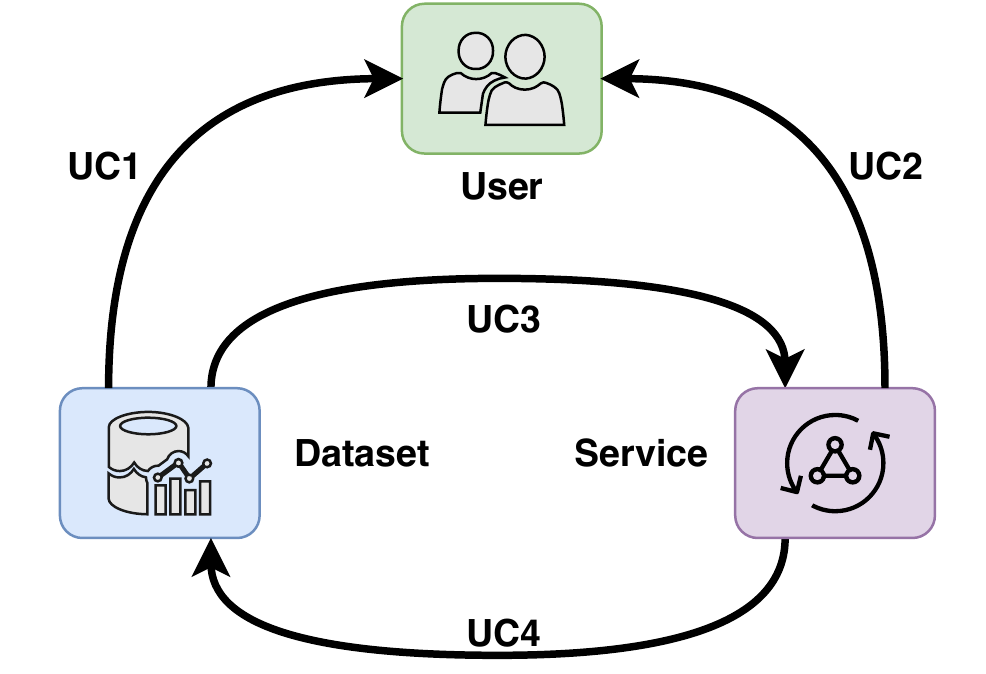}
   \caption{The tripartite relationship in a data market is spun between users, datasets and services. We identify four use cases for recommendations between these three identities, namely recommendation of datasets for users (UC1), recommendation of services for users (UC2), recommendation of datasets for services (UC3) and recommendation of services for datasets (UC4).}
	 \label{fig:dma-data}
\end{figure}

\para{Problem and objective of this work.} 
Recommender services thereby play a crucial role in data markets, since their suggestions allow to discover potential new combinations between users, datasets and services~\cite{damiani2015applying}. 
This results in a more complex tripartite relationship comprising users, datasets and services, as well as an increased number of use cases, in comparison with a traditional recommender setting. The tripartite structure and use cases are depicted in Figure~\ref{fig:dma-data}. 

However, most of the research in recommender systems is focused on settings consisting only of users and items, like recommending new movies to viewers. Hence, these settings can be categorized as bipartite relationships.
The work of~\cite{Godoy2016} points out the research need for recommendations in tripartite relationship scenarios such as the data markets scenario investigated in the work at hand. Another issue is the lack of an open dataset for the evaluation of tripartite recommendations in data markets. Therefore, we propose the use of the open Meta Kaggle dataset of the well-known data science portal Kaggle.

\para{Contributions and findings.}
The contributions of our work are two-fold:
\begin{itemize}
    \item We propose four use cases as well as a system architecture for recommendations in data markets (see Section~\ref{s:approach}).
    \item We provide evaluation results for a popularity-based as well as collaborative filtering-based algorithm for these four use cases using the open Meta Kaggle dataset (see Section~\ref{s:eval}).
\end{itemize}

\begin{figure*}[t!]
   \centering
	\includegraphics[width=0.85\textwidth]{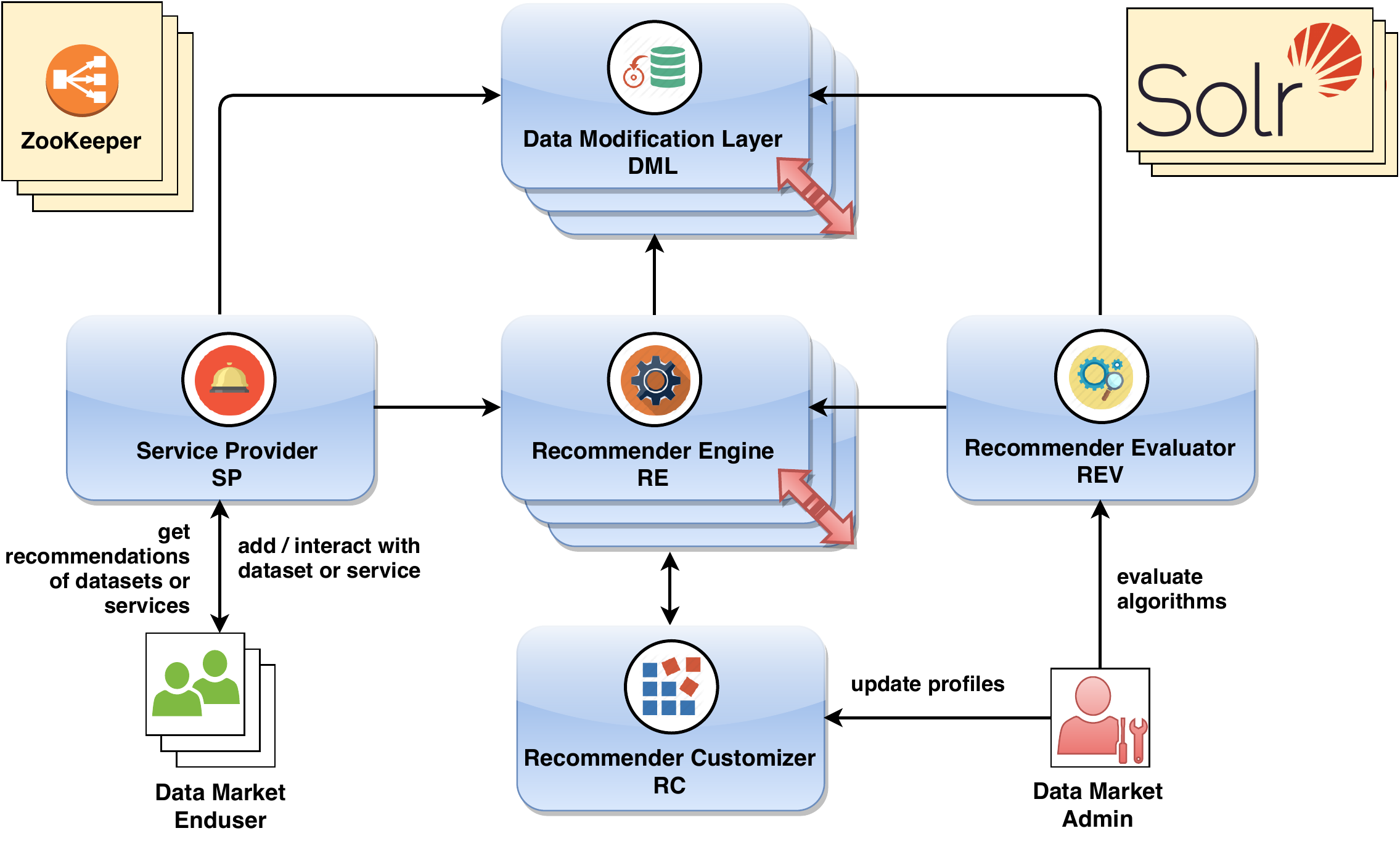}
   \caption{The system architecture of our recommender system for data markets is build upon ScaR as well as the open-source frameworks Apache Solr and ZooKeeper. The communication between the modules is handled via REST-based Web services.}
	 \label{fig:dma-rec}
\end{figure*}

Our results show that the recommendation accuracy strongly depends on the given use case. For example, in settings in which we have a limited set of candidate entities to recommend, already the simple popularity-based approach (recommending the most popular (MP) entities) provides good results. However, in more complex settings, where it is required to link services and datasets, a personalized approach such as collaborative filtering (CF) should be favored.

Taken together, our work contributes to the under-researched portfolio of recommender systems for data markets and thus, should be of interest for both researchers and practitioners in this area.

\section{Recommendations in Data Markets} \label{s:approach}
This section gives an a detailed overview of the four central data market use cases followed by the architecture of the proposed recommender system and all its components.

\subsection{Use Cases} \label{s:use}
As depicted in Figure~\ref{fig:dma-data}, data markets create a tripartite relationship between their entities users, datasets and services, thus leading to more complex recommendation problems. We identify four use cases for recommendations in the setting of data markets, investigated in more detail in the remainder of this subsection. 

\para{UC1: Recommendation of datasets for users.}
In the first use case, we recommend datasets to users. Thus, this one reflects a rather classic item2user recommendation problem, in which we analyze past user interactions between the target user and datasets (e.\,g., clicks or purchases) in order to recommend other datasets that could be interesting for the user (e.\,g., by using CF).

\para{UC2: Recommendation of services for users.}
The second use case also reflects a classic item2user recommendation problem but this time we aim to recommend services for users of the data market. Since typically there are more services than datasets available in a data market (see Section~\ref{s:data}), the set of potential candidate services is also larger, which makes this recommendation problem potentially harder than the one of UC1.

\para{UC3: Recommendation of datasets for services.}
UC3 reflects a more complex recommendation problem, in which we aim to recommend datasets for services. As both entities are now item types, we do no longer have classic user interactions for CF as we have in UC1 and UC2. To overcome this, we could establish an indirect connection between a dataset and a service when a user has interacted with both, the dataset and the service (see Section~\ref{s:data}). 

\para{UC4: Recommendation of services for datasets.}
In the fourth and final use case, we recommend services for datasets. As mentioned in UC2, we typically have more services than datasets available in a data market, which makes this use cases more complex than UC3, where the set of candidate entities to recommend is smaller. Furthermore, in UC4, we want to link services and datasets, where we do not have direct user interactions available. Thus, we believe that this use case is the most complex one and therefore, we also expect the lowest recommendation accuracy for this one (see Section~\ref{s:results}). 

\subsection{System Architecture} \label{s:framework}
The design of the system architecture of our recommender system for data markets is centred upon the scalable recommendation framework ScaR\footnote{\url{http://scar.know-center.tugraz.at/}}~\cite{lacic2015scar,lacic2014towards}. In Figure~\ref{fig:dma-rec}, we illustrate our main modules and how they interact with each other as well as with users and administrators of a data market. Apache ZooKeeper\footnote{\url{https://zookeeper.apache.org/}} is used for handling the communication between the modules and for load balancing (e.g., deploying multiple instances of a module).

\para{Service Provider (SP).}
The SP acts as a proxy for data markets to interact with the recommender system. It provides REST-based Web services to enable users to query recommendations of datasets and services, and to add new data (e.\,g., user interactions, datasets or services) to the system.

\para{Data Modification Layer (DML) \& Apache Solr.}
The DML encapsulates all database-related CRUD operations (i.\,e., create, retrieve, update, delete) in one module and thus, enables easy access to the underlying data backend. As shown in Figure~\ref{fig:dma-rec}, we utilize the high-performance search engine Apache Solr\footnote{\url{http://lucene.apache.org/solr/}}. This data backend solution not only guarantees scalability and (near) real-time recommendations but also the support of multiple entities like the users, datasets and services we encounter here.

\para{Recommender Engine (RE).}
The RE is the main module of our recommender system for data markets as it is responsible for calculating recommendations. Here, we make use of Apache Solr's build-in data structures for efficient similarity calculations. Currently, we focus on popularity and CF-based recommendation algorithms, but the RE module could be easily extended with further algorithms as well (e.\,g., content-based filtering).

\para{Recommender Customizer (RC).}
The RC is used to change the parameters (e.\,g., the number of recommended entities $k$) of the recommendation approaches on the fly. Thus, it holds a so-called recommendation profile for each approach, accessible and changeable by the data market administrator. These changes are then broadcast to the RE to be aware of how a specific approach should be executed.

\para{Recommender Evaluator (REV).}
The REV is responsible for evaluating the recommendation algorithms implemented in the RE. Hence, it can be executed to perform an offline evaluation with training/test set splits (see Section~\ref{s:metric}). In the future, it will also be possible to conduct online evaluations in data markets via A/B-tests.

\section{Evaluation} \label{s:eval}
In this section, we present our evaluation study, in which we compare popularity-based with CF-based recommendations for all four use cases defined in Section~\ref{s:use}.

\subsection{Data} \label{s:data}
For our evaluation, we use the open Meta Kaggle dataset\footnote{\url{https://www.kaggle.com/kaggle/meta-kaggle}} (2017-11-15) of the well-known Kaggle data science portal in order to simulate a real-world data market. Here, we have 6,108 users and 45 datasets that are connected via 2,926 user/dataset interactions, where an interaction is given by a user writing about a dataset in a discussion thread. Furthermore, we have 3,334 services that are connected to the 6,108 users via 18,593 user interactions. These interactions are created by users voting for a service. 

Finally, we establish a collaboration network between datasets and services (see e.g.,~\cite{hasani2018consensus}). Thus, we create a link between a dataset and a service when a user has interacted with both, the dataset and the service, which leads to 95,249 interactions. The full statistics of our dataset are summarized in Table~\ref{tab:datasets}.

\begin{table}[t!]
  \setlength{\tabcolsep}{16.0pt}	
  \centering	
    \begin{tabular}{l|c}
    \specialrule{.2em}{.1em}{.1em}
		Feature             & \#                         \\\hline
		Number of users	    & 6,108			                    \\
		Number of datasets  & 45			                    \\		
		Number of services	& 3,334		                    \\\hline
		Number of user/dataset interactions 		& 2,962					\\
		Number of user/service interactions	& 18,593			 		\\
		Number of dataset/service interactions & 95,249		    \\
		\specialrule{.2em}{.1em}{.1em}								
    \end{tabular}
	  \caption{Statistics of our dataset, which was collected from the Meta Kaggle platform in order to simulate the entities and interactions in a real-world data market. As the number of datasets is much smaller than the number of services, we also expect better recommendation accuracy results for UC1 and UC3 than for UC2 and UC4.}
  \label{tab:datasets}
\end{table}

\subsection{Evaluation Method} \label{s:metric}
In this section, we describe the evaluation protocol, recommendation algorithms and evaluation metrics used for our study.

\para{Evaluation protocol.}
For measuring the recommendation quality in the settings of the four use cases UC1 - UC4, we follow common practice in the area of recommender systems and split our Meta Kaggle dataset into training and test splits as suggested by~\cite{herlocker2004evaluating}.

Specifically, we extract all entities with at least eleven interactions from whom we withhold ten interactions for the test set and use the rest for training\footnote{We only evaluate users with a minimum of eleven interactions to ensure that we have at least one interactions for training when using ten interactions for testing.}~\cite{parra2009collaborative}. Thus, for UC1, this results into 17 users for whom we recommend datasets; for UC2, this results into 184 users for whom we recommend services; for UC3, this results into 2,338 services for whom we recommend datasets; for UC4, this results into 44 datasets for whom we recommend services.

\para{Recommendation algorithms.}
We evaluate our four uses cases for recommendations in data markets with two algorithms, namely most popular (MP) and (ii) collaborative filtering (CF). The recommendations are calculated and evaluated using the recommender system presented in Section~\ref{s:framework}.

MP is a non-personalized algorithm and is especially useful for new entities in a data market without any interactions so far, commonly referred as cold-start entities~\cite{schein2002methods}. This approach recommends datasets or services, which are weighted and ranked by the number of interactions. As mentioned, the MP approach is non-personalized and thus, each entity will receive the same recommendations.

CF algorithms~\cite{ricci2011introduction} analyze the interactions between users and entities, e.\,g., datasets and services alike. In CF methods two users are treated as similar if they have interacted with similar entities in the past. Hence, entities a similar user has interacted with in the past are candidates to recommend to a target user, who has not interacted with those entities yet. In the case of data markets, we do not only have interactions between users and entities but also between entities themselves when we consider UC3 and UC4. Here, we realize the CF approach in a similar way but instead of calculating user similarities, we calculate similarities between datasets and services, respectively.

\para{Evaluation metrics.} 
For measuring the accuracy of the recommendations in data markets, we use a rich set of metrics, namely Precision (P@$k$), F1-score (F1@$k$), Recall (R@$k$), Mean Reciprocal Rank (MRR@$k$), Mean Average Precision (MAP@$k$) and normalized Discounted Cumulative Gain (nDCG@$k$)~\cite{jarvelin2002cumulated}.

We report these metrics for different numbers of recommended entities (= $k$), i.\,e., P@1 for $k$ = 1, F1@5 for $k$ = 5\footnote{For 10 recommended entities, Precision typically reaches its highest value for $k$ = 1 and F1 for $k$ = 5.}, R@10 for $k$ = 10, MRR@10 for $k$ = 10, MAP@10 for $k$ = 10 and nDCG@10 for $k$ = 10. Please note that we set the maximum number of $k$ to 10, which is a common value for the evaluation of recommender systems~\cite{said2014comparative}.

\subsection{Results} \label{s:results}
In this section, we present the results of our evaluation with respect to our four use cases. Table~\ref{tab:results} holds the resulting numbers achieved in our experiments.

\begin{table}[t!]
  \setlength{\tabcolsep}{1.9pt}	
  \centering
    \begin{tabular}{l|cccccc}
    \specialrule{.2em}{.1em}{.1em}
		Approach  	& P@1	& F1@5		& R@10		& MRR@10	& MAP@10	& nDCG@10 \\\hline
		UC1: MP	  & \textbf{0.823}		& \textbf{0.470}	& \textbf{0.717}			& \textbf{0.217}	& \textbf{0.597}		& \textbf{0.729}	\\
		UC1: CF	& 0.705	& 0.431	& 0.611		& 0.192	& 0.484	& 0.635		\\\hline
		
		UC2: MP	 & 0.103 	& 0.050 	& 0.066		& 0.023	& 0.026	& 0.072		\\
		UC2: CF	 & \textbf{0.137} 	& \textbf{0.086}    & \textbf{0.114}		& \textbf{0.037}	& \textbf{0.054}	& \textbf{0.121}		\\\hline
		
		UC3: MP	 & 1.000	& 0.411		& 0.707	  	& 0.232		& 0.580		& 0.750	\\
		UC3: CF		  & \textbf{1.000}	& \textbf{0.636}	& \textbf{0.934}		& \textbf{0.281}		& \textbf{0.925}		& \textbf{0.948}						\\\hline
		
		UC4: MP			& 0.000	  & 0.000	& 0.000	& 0.000		& 0.000		& 0.000	\\
		UC4: CF		 & \textbf{0.022}	& \textbf{0.006}	& \textbf{0.006}	 	& \textbf{0.003}		& \textbf{0.004}		& \textbf{0.009}	 					\\
		
		\specialrule{.2em}{.1em}{.1em}								
    \end{tabular}
    \caption{Evaluation results of our four use cases for recommendations in data markets. While we get the best recommendation accuracy results for the unpersonalized MP algorithm in UC1, the personalized CF approach provides the best results for the more complex UC2, UC3 and UC4. The poor results for UC4 indicate that we need more sophisticated algorithms than MP and CF in this setting. Please note that bold numbers indicate the best results for a use case.}
  \label{tab:results}
\end{table}

\para{UC1: Recommendation of datasets for users.}
This use case reflects the least complex one as we recommend from a quite limited set of candidate entities (i.e., 45 datasets) with a small number of connections to the target entities (i.\,e., 2,962 user interactions). This is also reflected in the recommendation accuracy results presented in Table~\ref{tab:results} as the unpersonalized MP approach provides better results than the personalized CF one. This behavior of MP outperforming CF can only be observed in this use case, which shows that personalized approaches are not always necessary.

\para{UC2: Recommendation of services for users.}
When recommending services for users, we face a more complex problem since we have a much larger set of candidate entities (i.e., 3,334 services). Thus, the accuracy results in UC2 are much lower than the ones in UC1. Furthermore, in this case, the CF approach, which analyzes the 18,593 interactions between users and services in a personalized manner, provides better results than MP.

\para{UC3: Recommendation of datasets for services.}
Similar to UC1, in UC3, we also recommend datasets but this time for services instead of users. For this use case, we also have a large set of 95,249 interactions between datasets and services available, leading to the overall best results for CF across all four use cases. Interestingly, both MP and CF provide a perfect score for P@1 of 1.000, which indicates that both algorithms rank a highly-connected dataset on the first position that is relevant for all 2,338 evaluated services.

\para{UC4: Recommendation of services for datasets.}
UC4 reflects the most complex of our use cases since we have a large set of 3,334 candidates services available, which are linked via 95,249 interactions to a small set of 44 datasets being the evaluated entities. This is reflected in the results shown in Table~\ref{tab:results} as both algorithms, MP and CF, provide the worst results across all use cases. Here, the unpersonalized MP approach even reaches a recommendation accuracy of 0.000 for all metrics, thus not recommending a single relevant service.

\subsection{Discussion} \label{s:disc}
Our evaluation results show that there is no one-size-fits-all solution for recommendations in data markets. One particular finding of us is, that in cases having a limited set of candidate entities available like in UC1, popularity-based methods such as MP provide good results. Another finding is that personalized methods such as CF should be favored when the use cases get more complex, for example if we have a larger set of candidate entities as it is the case in UC2. The same holds for the recommendations of entities to other entities, like datasets to services in UC3. 

However, our results also show that both MP and CF provide poor results for UC4 being the most complex use case. For such a setting, we need more sophisticated methods that incorporate also other data sources, e.\,g., content-based filtering (CBF) approaches~\cite{lops2011content}. For overcoming sparsity problems, these approaches could also be combined with word embeddings~\cite{kenter2015short,mikolov2013word2vec}.

\section{Conclusion and Future Work} \label{s:conc}
In this paper, we presented our initial steps for providing and evaluating recommendations in data markets. Therefore, we first provided four potential use cases, which included recommendation of datasets for users (UC1), recommendation of services for users (UC2), recommendation of datasets for services (UC3), and recommendation of services for datasets (UC4). Then, we proposed a system architecture for a recommender system for data markets based on the scalable recommendation framework ScaR.

Finally, we provided an evaluation of these four uses using the Meta Kaggle dataset and our proposed recommender system. Here, we find that the unpersonalized most popular approach (MP) provides the best results for UC1 and the personalized collaborative filtering approach (CF) provides the best results for the more complex use cases UC2, UC3 and UC4.

\para{Limitations and future Work.}
One limitation of our evaluation is that we have simulated a real-world data market using the Meta Kaggle dataset. Although, this dataset provides all relevant entities of data markets (i.\,e., users, datasets and services), we plan to also conduct evaluation studies in real-world data markets such as the ones created in the DMA and AI4EU initiatives.

Furthermore, so far, we have only evaluated the two algorithms MP and CF. Thus, we also plan to extend our study with more recommendation approaches such as content-based filtering (see Section~\ref{s:disc}).

\para{Acknowledgments.}
This work was supported by the Know-Center GmbH, the FFG flagship project Data Market Austria (DMA) and the H2020 project AI4EU (GA: 825619). The Know-Center GmbH is funded within the Austrian COMET Program - Competence Centers for Excellent Technologies - under the auspices of the Austrian Ministry of Transport, Innovation and Technology, the Austrian Ministry of Economics and Labor and by the State of Styria. COMET is managed by the Austrian Research Promotion Agency (FFG).

\bibliographystyle{ACM-Reference-Format}
\input{document.bbl}

\end{document}

%% file: document.bbl

%% file: document.bbl
\begin{thebibliography}{00}


\ifx \showCODEN    \undefined \def \showCODEN     #1{\unskip}     \fi
\ifx \showDOI      \undefined \def \showDOI       #1{{\tt DOI:}\penalty0{#1}\ }
  \fi
\ifx \showISBNx    \undefined \def \showISBNx     #1{\unskip}     \fi
\ifx \showISBNxiii \undefined \def \showISBNxiii  #1{\unskip}     \fi
\ifx \showISSN     \undefined \def \showISSN      #1{\unskip}     \fi
\ifx \showLCCN     \undefined \def \showLCCN      #1{\unskip}     \fi
\ifx \shownote     \undefined \def \shownote      #1{#1}          \fi
\ifx \showarticletitle \undefined \def \showarticletitle #1{#1}   \fi
\ifx \showURL      \undefined \def \showURL       #1{#1}          \fi
\providecommand\bibfield[2]{#2}
\providecommand\bibinfo[2]{#2}
\providecommand\natexlab[1]{#1}
\providecommand\showeprint[2][]{arXiv:#2}

\bibitem[\protect\citeauthoryear{Cavanillas, Curry, and Wahlster}{Cavanillas
  et~al\mbox{.}}{2016}]%
        {Cavanillas2016}
\bibfield{editor}{\bibinfo{person}{Jos\'{e}~Mar\'{i}a Cavanillas},
  \bibinfo{person}{Edward Curry}, {and} \bibinfo{person}{Wolfgang Wahlster}}
  (Eds.). \bibinfo{year}{2016}\natexlab{}.
\newblock \bibinfo{booktitle}{{\em New Horizons for a Data-Driven Economy}}.
\newblock \bibinfo{publisher}{Springer}, \bibinfo{address}{Cham, Switzerland}.
\newblock
\showISBNx{978-3-319-21569-3}
\showDOI{%
\url{http://dx.doi.org/10.1007/978-3-319-21569-3}}


\bibitem[\protect\citeauthoryear{Curry}{Curry}{2016}]%
        {curry2016big}
\bibfield{author}{\bibinfo{person}{Edward Curry}.}
  \bibinfo{year}{2016}\natexlab{}.
\newblock \showarticletitle{The big data value chain: definitions, concepts,
  and theoretical approaches}.
\newblock In \bibinfo{booktitle}{{\em New horizons for a data-driven economy}}.
  \bibinfo{publisher}{Springer, Cham}, \bibinfo{pages}{29--37}.
\newblock


\bibitem[\protect\citeauthoryear{Damiani, Ceravolo, Frati, Bellandi, Maier,
  Seeber, and Waldhart}{Damiani et~al\mbox{.}}{2015}]%
        {damiani2015applying}
\bibfield{author}{\bibinfo{person}{Ernesto Damiani}, \bibinfo{person}{Paolo
  Ceravolo}, \bibinfo{person}{Fulvio Frati}, \bibinfo{person}{Valerio
  Bellandi}, \bibinfo{person}{Ronald Maier}, \bibinfo{person}{Isabella Seeber},
  {and} \bibinfo{person}{Gabriela Waldhart}.} \bibinfo{year}{2015}\natexlab{}.
\newblock \showarticletitle{Applying recommender systems in collaboration
  environments}.
\newblock \bibinfo{journal}{{\em Computers in Human Behavior\/}}
  \bibinfo{volume}{51} (\bibinfo{year}{2015}), \bibinfo{pages}{1124--1133}.
\newblock


\bibitem[\protect\citeauthoryear{Godoy and Corbellini}{Godoy and
  Corbellini}{2016}]%
        {Godoy2016}
\bibfield{author}{\bibinfo{person}{Daniela Godoy} {and}
  \bibinfo{person}{Alejandro Corbellini}.} \bibinfo{year}{2016}\natexlab{}.
\newblock \showarticletitle{Folksonomy-Based Recommender Systems: A
  State-of-the-Art Review}.
\newblock \bibinfo{journal}{{\em International Journal of Intelligent
  Systems\/}} \bibinfo{volume}{31}, \bibinfo{number}{4} (\bibinfo{year}{2016}),
  \bibinfo{pages}{314--346}.
\newblock
\showDOI{%
\url{http://dx.doi.org/10.1002/int.21753}}
\showeprint{https://onlinelibrary.wiley.com/doi/pdf/10.1002/int.21753}


\bibitem[\protect\citeauthoryear{Hasani-Mavriqi, Kowald, Helic, and
  Lex}{Hasani-Mavriqi et~al\mbox{.}}{2018}]%
        {hasani2018consensus}
\bibfield{author}{\bibinfo{person}{Ilire Hasani-Mavriqi},
  \bibinfo{person}{Dominik Kowald}, \bibinfo{person}{Denis Helic}, {and}
  \bibinfo{person}{Elisabeth Lex}.} \bibinfo{year}{2018}\natexlab{}.
\newblock \showarticletitle{Consensus dynamics in online collaboration
  systems}.
\newblock \bibinfo{journal}{{\em Computational Social Networks\/}}
  \bibinfo{volume}{5}, \bibinfo{number}{1} (\bibinfo{year}{2018}),
  \bibinfo{pages}{2}.
\newblock


\bibitem[\protect\citeauthoryear{Herlocker, Konstan, Terveen, and
  Riedl}{Herlocker et~al\mbox{.}}{2004}]%
        {herlocker2004evaluating}
\bibfield{author}{\bibinfo{person}{Jonathan~L Herlocker},
  \bibinfo{person}{Joseph~A Konstan}, \bibinfo{person}{Loren~G Terveen}, {and}
  \bibinfo{person}{John~T Riedl}.} \bibinfo{year}{2004}\natexlab{}.
\newblock \showarticletitle{Evaluating collaborative filtering recommender
  systems}.
\newblock \bibinfo{journal}{{\em ACM Transactions on Information Systems
  (TOIS)\/}} \bibinfo{volume}{22}, \bibinfo{number}{1} (\bibinfo{year}{2004}),
  \bibinfo{pages}{5--53}.
\newblock


\bibitem[\protect\citeauthoryear{J{\"a}rvelin and
  Kek{\"a}l{\"a}inen}{J{\"a}rvelin and Kek{\"a}l{\"a}inen}{2002}]%
        {jarvelin2002cumulated}
\bibfield{author}{\bibinfo{person}{Kalervo J{\"a}rvelin} {and}
  \bibinfo{person}{Jaana Kek{\"a}l{\"a}inen}.} \bibinfo{year}{2002}\natexlab{}.
\newblock \showarticletitle{Cumulated gain-based evaluation of IR techniques}.
\newblock \bibinfo{journal}{{\em ACM Transactions on Information Systems
  (TOIS)\/}} \bibinfo{volume}{20}, \bibinfo{number}{4} (\bibinfo{year}{2002}),
  \bibinfo{pages}{422--446}.
\newblock


\bibitem[\protect\citeauthoryear{Kenter and De~Rijke}{Kenter and
  De~Rijke}{2015}]%
        {kenter2015short}
\bibfield{author}{\bibinfo{person}{Tom Kenter} {and} \bibinfo{person}{Maarten
  De~Rijke}.} \bibinfo{year}{2015}\natexlab{}.
\newblock \showarticletitle{Short text similarity with word embeddings}. In
  \bibinfo{booktitle}{{\em Proc. of CIKM'15}}. ACM,
  \bibinfo{pages}{1411--1420}.
\newblock


\bibitem[\protect\citeauthoryear{Lacic, Kowald, Parra, Kahr, and
  Trattner}{Lacic et~al\mbox{.}}{2014}]%
        {lacic2014towards}
\bibfield{author}{\bibinfo{person}{Emanuel Lacic}, \bibinfo{person}{Dominik
  Kowald}, \bibinfo{person}{Denis Parra}, \bibinfo{person}{Martin Kahr}, {and}
  \bibinfo{person}{Christoph Trattner}.} \bibinfo{year}{2014}\natexlab{}.
\newblock \showarticletitle{Towards a scalable social recommender engine for
  online marketplaces: The case of apache solr}. In \bibinfo{booktitle}{{\em
  Proceedings of the 23rd International Conference on World Wide Web}}. ACM,
  \bibinfo{pages}{817--822}.
\newblock


\bibitem[\protect\citeauthoryear{Lacic, Traub, Kowald, and Lex}{Lacic
  et~al\mbox{.}}{2015}]%
        {lacic2015scar}
\bibfield{author}{\bibinfo{person}{Emanuel Lacic}, \bibinfo{person}{Matthias
  Traub}, \bibinfo{person}{Dominik Kowald}, {and} \bibinfo{person}{Elisabeth
  Lex}.} \bibinfo{year}{2015}\natexlab{}.
\newblock \showarticletitle{ScaR: Towards a Real-Time Recommender Framework
  Following the Microservices Architecture}. In \bibinfo{booktitle}{{\em
  Proceedings of LSRS2015 Workshop at RecSys 2015}}. .
\newblock


\bibitem[\protect\citeauthoryear{Lops, De~Gemmis, and Semeraro}{Lops
  et~al\mbox{.}}{2011}]%
        {lops2011content}
\bibfield{author}{\bibinfo{person}{Pasquale Lops}, \bibinfo{person}{Marco
  De~Gemmis}, {and} \bibinfo{person}{Giovanni Semeraro}.}
  \bibinfo{year}{2011}\natexlab{}.
\newblock \showarticletitle{Content-based recommender systems: State of the art
  and trends}.
\newblock In \bibinfo{booktitle}{{\em Recommender systems handbook}}.
  \bibinfo{publisher}{Springer}, \bibinfo{pages}{73--105}.
\newblock


\bibitem[\protect\citeauthoryear{{Mikolov}, {Chen}, {Corrado}, and
  {Dean}}{{Mikolov} et~al\mbox{.}}{2013}]%
        {mikolov2013word2vec}
\bibfield{author}{\bibinfo{person}{Tomas {Mikolov}}, \bibinfo{person}{Kai
  {Chen}}, \bibinfo{person}{Greg {Corrado}}, {and} \bibinfo{person}{Jeffrey
  {Dean}}.} \bibinfo{year}{2013}\natexlab{}.
\newblock \showarticletitle{{Efficient Estimation of Word Representations in
  Vector Space}}.
\newblock \bibinfo{journal}{{\em arXiv e-prints\/}} (\bibinfo{date}{Jan}
  \bibinfo{year}{2013}), \bibinfo{pages}{arXiv:1301.3781}.
\newblock
\showeprint[arxiv]{cs.CL/1301.3781}


\bibitem[\protect\citeauthoryear{Parra and Brusilovsky}{Parra and
  Brusilovsky}{2009}]%
        {parra2009collaborative}
\bibfield{author}{\bibinfo{person}{Denis Parra} {and} \bibinfo{person}{Peter
  Brusilovsky}.} \bibinfo{year}{2009}\natexlab{}.
\newblock \showarticletitle{Collaborative filtering for social tagging systems:
  an experiment with CiteULike}. In \bibinfo{booktitle}{{\em Proceedings of the
  third ACM conference on Recommender systems}}. ACM,
  \bibinfo{pages}{237--240}.
\newblock


\bibitem[\protect\citeauthoryear{Ricci, Rokach, and Shapira}{Ricci
  et~al\mbox{.}}{2011}]%
        {ricci2011introduction}
\bibfield{author}{\bibinfo{person}{Francesco Ricci}, \bibinfo{person}{Lior
  Rokach}, {and} \bibinfo{person}{Bracha Shapira}.}
  \bibinfo{year}{2011}\natexlab{}.
\newblock \bibinfo{booktitle}{{\em Introduction to recommender systems
  handbook}}.
\newblock \bibinfo{publisher}{Springer}.
\newblock


\bibitem[\protect\citeauthoryear{Said and Bellog{\'\i}n}{Said and
  Bellog{\'\i}n}{2014}]%
        {said2014comparative}
\bibfield{author}{\bibinfo{person}{Alan Said} {and} \bibinfo{person}{Alejandro
  Bellog{\'\i}n}.} \bibinfo{year}{2014}\natexlab{}.
\newblock \showarticletitle{Comparative recommender system evaluation:
  benchmarking recommendation frameworks}. In \bibinfo{booktitle}{{\em
  Proceedings of the 8th ACM Conference on Recommender systems}}. ACM,
  \bibinfo{pages}{129--136}.
\newblock


\bibitem[\protect\citeauthoryear{Schein, Popescul, Ungar, and Pennock}{Schein
  et~al\mbox{.}}{2002}]%
        {schein2002methods}
\bibfield{author}{\bibinfo{person}{Andrew~I Schein},
  \bibinfo{person}{Alexandrin Popescul}, \bibinfo{person}{Lyle~H Ungar}, {and}
  \bibinfo{person}{David~M Pennock}.} \bibinfo{year}{2002}\natexlab{}.
\newblock \showarticletitle{Methods and metrics for cold-start
  recommendations}. In \bibinfo{booktitle}{{\em Proceedings of the 25th annual
  international ACM SIGIR conference on Research and development in information
  retrieval}}. ACM, \bibinfo{pages}{253--260}.
\newblock


\bibitem[\protect\citeauthoryear{Traub, Gursch, Lex, and Kern}{Traub
  et~al\mbox{.}}{2017}]%
        {traub2017}
\bibfield{author}{\bibinfo{person}{Matthias Traub}, \bibinfo{person}{Heimo
  Gursch}, \bibinfo{person}{Elisabeth Lex}, {and} \bibinfo{person}{Roman
  Kern}.} \bibinfo{year}{2017}\natexlab{}.
\newblock \showarticletitle{Data Market Austria} {\em
  (\bibinfo{series}{Institute of Systems Sciences, Innovation and
  Sustainability Reports})}, \bibfield{editor}{\bibinfo{person}{Romana Rauter},
  \bibinfo{person}{Martina Zimek}, \bibinfo{person}{Aisma~Linda Kiesnere},
  {and} \bibinfo{person}{Rupert~J. Baumgartner}} (Eds.).
  \bibinfo{publisher}{Institute of Systems Sciences, Innovation and
  Sustainability, University of Graz}, \bibinfo{pages}{353--363}.
\newblock
\showISSN{2308-1767}


\end{thebibliography}
